# Properties of modified periodic one-dimensional hopping model

Yunxin Zhang[*]

*Shanghai Key Laboratory for Contemporary Applied Mathematics,*

*Centre for Computational System Biology,*

*School of Mathematical Sciences, Fudan University, Shanghai 200433, China.*

(Dated: June 19, 2010)

One-dimensional hopping model is useful to describe the motion of microscopic particle in thermal noise environment, such as motor proteins. Recent experiments about the new generation of light-driven rotary molecular motors found that, the motor in state $i$ can jump forward to state $i+1$ or $i+2$, or backward to state $i-1$ or $i-2$ directly. In this paper, such modified periodic one-dimensional hopping model of arbitrary period $N$ is studied mathematically. The mean velocity, effective diffusion constant, and mean dwell time in one single cycle are obtained. Corresponding results are illustrated and verified by being applied to a type of synthetic rotary molecular motors.

## I. INTRODUCTION

Many physical [1, 2] and biochemical processes, for example, the motion of motor proteins kinesin, dynein, and myosin [3–9], can be well described by periodic one dimensional hopping models [10–15]. Mathematically, these models have been extensively studied, and the mean velocity, effective diffusion constant, and mean first passage time have been obtained [16–20]. In these models, the particle in state $i$ can jump forward to state $i+1$ with rate $u_i$, or backward to state $i-1$ with rate $w_i$. Where the forward and backward transition rates satisfy periodicity conditions $u_{N+i} = u_i$, $w_{N+i} = w_i$ with $N$ is the number of state in one cycle.

---

[*]Electronic address: `xyz@fudan.edu.cn`

However, recent experimental data of a type of synthetic light-driven rotary molecular motors, which was devised by Feringa and coworkers, found that, during the rotation of the molecular motor, the motor in state $i$ also can jump forward to state $i+2$, or backward to state $i-2$ directly [21–29]. In fact, it also has been found that, during the motion of many other molecular motors [30–35], there are usually more than two choices for the motor to jump out its present state. So it is valuable to study more general one-dimensional hopping models, in which the particle in state $i$ is allowed to jump to other states directly besides states $i-1$ and $i+1$. In this paper, one of the simple cases, which we called *modified* one-dimensional hopping model, during which, beside states $i-1$ and $i+1$, the particle in state $i$ also can jump forward directly to state $i+2$ with rate $u'_i$, or backward directly to state $i-2$ with rate $w'_i$, is theoretically analyzed. Using the similar idea as Derrida's [16], the mean velocity $V$ and effective diffusion constant $D$ are obtained. Meanwhile, by similar method as in [17], the mean dwell time in one single cycle is also obtained. The reason that we discuss this simple case here, is not only because it is theoretically tractable, but also the corresponding results can be verified by the experimental data [29].

The organization of this paper is as follows. The modified one-dimensional hopping model is described in the following Section and, then in Section **III**, the mean velocity and effective diffusion constant are obtained under the same assumptions used by Derrida [16]. In Section **IV**, we adopt the approach of Pury and Cáceres [17] for mean first passage times to calculate the mean dwell time in one single mechanochemical cycle. The results are illustrated and verified by being applied to Feringa's model for synthetic rotary molecular motor in Section **V**. Finally, a brief summary is given in Section **VI**.

## II. MODIFIED ONE-DIMENSIONAL HOPPING MODEL

Our modified one-dimensional hopping model of period $N$ is schematically depicted in Fig. 1: in which rate $u_i$ corresponds to a forward transition from state $i$ to state $i+1$, while rate $u'_i$ corresponds to a forward transition from state $i$ to state $i+2$; similarly, the rate $w_i$ corresponds to a backward transition from state $i$ to state $i-1$, while rate $w'_i$ corresponds to backward transition to state $i-2$. Periodicity requires $u_{i+N} = u_i, u'_{i+N} = u'_i, w_{i+N} = w_i, w'_{i+N} = w'_i$.

Following the analysis of [16, 36], if $\tilde{p}_i(t)$ is the probability of finding the particle in state $i$ at time $t$, then the master equation reads

$$\begin{aligned}\frac{d\tilde{p}_i(t)}{dt} &= (u'_{i-2}\tilde{p}_{i-2} + u_{i-1}\tilde{p}_{i-1} + w_{i+1}\tilde{p}_{i+1} + w'_{i+2}\tilde{p}_{i+2}) - (u_i + u'_i + w_i + w'_i)\tilde{p}_i \\ &=: \tilde{J}_{i-\frac{1}{2}} - \tilde{J}_{i+\frac{1}{2}},\end{aligned} \qquad (1)$$

where, for brevity we have introduced the flux current

$$\tilde{J}_{i-\frac{1}{2}} = (u'_{i-2}\tilde{p}_{i-2} + u_{i-1}\tilde{p}_{i-1} + u'_{i-1}\tilde{p}_{i-1}) - (w_i\tilde{p}_i + w'_i\tilde{p}_i + w'_{i+1}\tilde{p}_{i+1}). \qquad (2)$$

Let

$$\bar{p}_i = \sum_{k=-\infty}^{+\infty} \tilde{p}_{kN+i}, \qquad \bar{s}_i = \sum_{k=-\infty}^{+\infty} (kN+i)\tilde{p}_{kN+i}, \qquad (3)$$

then

$$\bar{p}_i = \bar{p}_{N+i}, \qquad \bar{s}_i = \bar{s}_{N+i}, \qquad \sum_{i=1}^{N} \bar{p}_i = \sum_{i=-\infty}^{+\infty} \tilde{p}_i = 1, \qquad (4)$$

and satisfy

$$\frac{d\bar{p}_i(t)}{dt} = (u'_{i-2}\bar{p}_{i-2} + u_{i-1}\bar{p}_{i-1} + w_{i+1}\bar{p}_{i+1} + w'_{i+2}\bar{p}_{i+2}) - (u_i + u'_i + w_i + w'_i)\bar{p}_i, \qquad (5)$$

$$\begin{aligned}\frac{d\bar{s}_i(t)}{dt} &= (u'_{i-2}\bar{s}_{i-2} + u_{i-1}\bar{s}_{i-1} + w_{i+1}\bar{s}_{i+1} + w'_{i+2}\bar{s}_{i+2}) - (u_i + u'_i + w_i + w'_i)\bar{s}_i \\ &\quad + 2u'_{i-2}\bar{p}_{i-2} + u_{i-1}\bar{p}_{i-1} - w_{i+1}\bar{p}_{i+1} - 2w'_{i+2}\bar{p}_{i+1}.\end{aligned} \qquad (6)$$

In the long time limit, we assume, following Derrida [16], that

$$\bar{p}_i(t) \to p_i, \quad \text{and} \quad \bar{s}_i(t) \to a_i t + b_i, \qquad (7)$$

where $p_i, a_i, b_i$ are constants to be determined. It can be easily verified that they must satisfy the relations

$$0 = (u'_{i-2}p_{i-2} + u_{i-1}p_{i-1} + w_{i+1}p_{i+1} + w'_{i+2}p_{i+2}) - (u_i + u'_i + w_i + w'_i)p_i, \qquad (8)$$

$$0 = (u'_{i-2}a_{i-2} + u_{i-1}a_{i-1} + w_{i+1}a_{i+1} + w'_{i+2}a_{i+2}) - (u_i + u'_i + w_i + w'_i)a_i, \qquad (9)$$

and

$$\begin{aligned}a_i &= (u'_{i-2}b_{i-2} + u_{i-1}b_{i-1} + w_{i+1}b_{i+1} + w'_{i+2}b_{i+2}) - (u_i + u'_i + w_i + w'_i)b_i \\ &\quad + 2u'_{i-2}p_{i-2} + u_{i-1}p_{i-1} - w_{i+1}p_{i+1} - 2w'_{i+2}p_{i+2}.\end{aligned} \qquad (10)$$

Analogously to the definition in (2), let us define

$$J_{i-\frac{1}{2}} = (u'_{i-2}p_{i-2} + u_{i-1}p_{i-1} + u'_{i-1}p_{i-1}) - (w_i p_i + w'_i p_i + w'_{i+1} p_{i+1}). \tag{11}$$

Then Eq. (8) implies

$$J_{i-\frac{1}{2}} =: J, \tag{12}$$

is constant for any $1 \leq i \leq N$. For the sake of convenience, we rewrite the linear algebraic equations (8) as

$$A\vec{p} = 0, \tag{13}$$

where $\vec{p} = (p_1, \cdots, p_N)^T$ are column vector, while

$$\begin{aligned}
& A_{i\,i-2} = u'_{i-2}, \quad A_{i\,i-1} = u_{i-1}, \quad A_{i\,i} = -(u''_i + w''_i), \\
& A_{i\,i+1} = w_{i+1}, \quad A_{i\,i+2} = w'_{i+2}, \quad \text{for } 3 \leq i \leq N-2; \\
& A_{1\,N-1} = u'_{N-1}, \quad A_{1\,N} = u_{N-1}, \quad A_{1\,1} = -(u''_1 + w''_1), \\
& A_{1\,2} = w_2, \quad A_{1\,3} = w'_3; \\
& A_{2\,N} = u'_N, \quad A_{2\,1} = u_1, \quad A_{2\,2} = -(u''_2 + w''_2), \\
& A_{2\,3} = w_3, \quad A_{2\,4} = w'_4; \\
& A_{N-1\,N-3} = u'_{N-3}, \quad A_{N-1\,N-1} = -(u''_{N-1} + w''_{N-1}), \\
& A_{N-1\,N-2} = u_{N-2}, \quad A_{N-1\,N} = w_N, \quad A_{N-1\,1} = w'_1; \\
& A_{N\,N-2} = u'_{N-2}, \quad A_{N\,N-1} = u_{N-1}, \quad A_{N\,N} = -(u''_N + w''_N), \\
& A_{N\,1} = w_1, \quad A_{N\,2} = w'_2; \\
& A_{i\,j} = 0, \quad \text{otherwise}.
\end{aligned} \tag{14}$$

where $u''_i := u_i + u'_i$, and $w''_i := w_i + w'_i$.

Usually the steady state probability $p_i$ for $1 \leq i \leq N$ can be determined by Eq. (8) or (13) and the normalization condition $\sum_{i=1}^{N} p_i = 1$. In our discussion, we always assume this fact holds, which implies that the characteristic space of eigenvalue 0 of matrix $A$ is one-dimensional (it is easy to see that $\det(A) = 0$, since $\sum_{i=1}^{N} A_{ij} = 0$ for any $1 \leq j \leq N$). So, from (8) and (9), one can see

$$a_i = \overline{V} p_i, \tag{15}$$

where $\overline{V}$ is another constant. Combining with the normalizing condition $\sum_{i=1}^{N} p_i = 1$ one easily finds

$$\begin{aligned}\overline{V} &= \sum_{i=1}^{N} a_i, \\ &= \sum_{i=1}^{N} \left(2u'_{i-2}p_{i-2} + u_{i-1}p_{i-1} - w_{i+1}p_{i+1} - 2w'_{i+2}p_{i+2}\right), \\ &= \sum_{i=1}^{N} J_{i-\frac{1}{2}} = NJ.\end{aligned} \qquad(16)$$

## III. MEAN VELOCITY AND EFFECTIVE DIFFUSION CONSTANT

Physically, the definitions of the mean velocity $V$ and the effective diffusion constant $D$ are (see [16, 19])

$$\begin{aligned}V &= \lim_{t\to\infty} \frac{d\langle x(t)\rangle}{dt}, \\ D &= \frac{1}{2}\lim_{t\to\infty} \frac{d}{dt}\left[\langle x^2(t)\rangle - \langle x(t)\rangle^2\right],\end{aligned} \qquad(17)$$

where $\langle x^k(t)\rangle \triangleq \sum_{i=-\infty}^{\infty} i^k \tilde{p}_i(t)$. Now we can easily shown that

$$\begin{aligned}V &= \lim_{t\to\infty} \frac{d\langle x(t)\rangle}{dt} = \lim_{t\to\infty} \frac{d}{dt} \sum_{i=1}^{N} \sum_{k=-\infty}^{\infty} (kN+i)\tilde{p}_{kN+i} = \lim_{t\to\infty} \sum_{i=1}^{N} \frac{d\bar{s}_i}{dt}, \\ &= \lim_{t\to\infty} \sum_{i=1}^{N} (2u'_{i-2}\bar{p}_{i-2} + u_{i-1}\bar{p}_{i-1} - w_{i+1}\bar{p}_{i+1} - 2w'_{i+2}\bar{p}_{i+2}), \\ &= \sum_{i=1}^{N} (2u'_{i-2}p_{i-2} + u_{i-1}p_{i-1} - w_{i+1}p_{i+1} - 2w'_{i+2}p_{i+2}), \\ &= NJ = \overline{V},\end{aligned} \qquad(18)$$

while one can also obtain

$$\begin{aligned}\frac{d\langle x^2(t)\rangle}{dt} &= \sum_{i=1}^{N} \sum_{k=-\infty}^{\infty} (kN+i)^2[-(u_i + u'_i + w_i + w'_i)\tilde{p}_{kN+i} \\ &\quad + u'_{i-2}\tilde{p}_{kN+i-2} + u_{i-1}\tilde{p}_{kN+i-1} + w_{i+1}\tilde{p}_{kN+i+1} + w'_{i+2}\tilde{p}_{kN+i+2}], \\ &= \sum_{i=1}^{N}[(4u'_i + 2u_i - 2w_i - 4w'_i)\bar{s}_i + (4u'_i + u_i + w_i + 4w'_i)\bar{p}_i].\end{aligned} \qquad(19)$$

Therefore, combining Eqs. (7) (17) (18) and (19), we may conclude,

$$D = \sum_{i=1}^{N}[(2u'_i + u_i - w_i - 2w'_i)(a_i t + b_i) + \frac{1}{2}(4u'_i + u_i + w_i + 4w'_i)p_i]$$
$$- V \sum_{i=1}^{N}(a_i t + b_i), \tag{20}$$
$$= \sum_{i=1}^{N}[(2u'_i + u_i - w_i - 2w'_i)b_i + \frac{1}{2}(4u'_i + u_i + w_i + 4w'_i)p_i] - V\sum_{i=1}^{N} b_i,$$

where $V = \overline{V}, p_i$ can be obtained with the aid of (13) and (16) (18), while $\vec{b} = (b_1, \cdots, b_N)^T$ can be found by (10), which leads to

$$A\vec{b} = \vec{a} - G\vec{p} = (VI - G)\vec{p}, \tag{21}$$

where $I = \text{diag}(1, \cdots, 1)$ is unit matrix, and $\vec{a} = (a_1, \cdots, a_N)^T$ is a column vector, while

$$G_{i\,i-2} = 2u'_{i-2},\ G_{i\,i-1} = u_{i-1},\ G_{i\,i+1} = -w_{i+1},$$
$$G_{i\,i+2} = -2w'_{i+2},\ \text{ for } 3 \leq i \leq N-2;$$
$$G_{1\,N-1} = 2u'_{N-1},\ G_{1\,N} = u_N,\ G_{1\,2} = -w_2,\ G_{1\,3} = -2w'_3;$$
$$G_{2\,N} = 2u'_N,\ G_{2\,1} = u_1,\ G_{2\,3} = -w_3,\ G_{2\,4} = -2w'_4;$$
$$G_{N-1\,N-3} = 2u'_{N-3},\ G_{N-1\,N-2} = u_{N-2},\ G_{N-1\,N} = -w_N,\ G_{N-1\,1} = -2w'_1;$$
$$G_{N\,N-2} = 2u'_{N-2},\ G_{N\,N-1} = u_{N-1},\ G_{N\,1} = -w_1,\ G_{N\,2} = -2w'_2;$$
$$G_{ij} = 0,\ \text{ otherwise.} \tag{22}$$

Since $\det(A) = 0$, so $\vec{b}$ cannot be uniquely determined by Eq. (21). However, the value of the effective diffusion constant $D$ is uniquely determined by (20) and (21). In fact, if $\vec{b}_1, \vec{b}_2$ are two solutions of (21), then $\vec{b}_0 = \vec{b}_1 - \vec{b}_2$ satisfies $A\vec{b}_0 = 0$. Since the characteristic space of eigenvalue 0 of $A$ is one-dimensional, so, from (13), one finds that $\vec{b}_0 = c\vec{p}$ with constant $c = \left(\sum_{k=1}^{N} b_{0k}\right)$, i.e., $b_{0i} = \left(\sum_{k=1}^{N} b_{0k}\right)p_i$.

Thus,

$$
\begin{aligned}
D(\vec{b}_1) - D(\vec{b}_2) &= \sum_{i=1}^{N}(2u'_i + u_i - w_i - 2w'_i)b_{0i} - V\sum_{k=1}^{N} b_{0k}, \\
&= \sum_{i=1}^{N}(2u'_i + u_i - w_i - 2w'_i)b_{0i} \\
&\quad - \left(\sum_{i=1}^{N}(2u'_{i-2}p_{i-2} + u_{i-1}p_{i-1} - w_{i+1}p_{i+1} - 2w'_{i+2}p_{i+2})\right)\left(\sum_{k=1}^{N} b_{0k}\right), \\
&= \sum_{i=1}^{N}(2u'_i + u_i - w_i - 2w'_i)b_{0i} - \left(\sum_{i=1}^{N}(2u'_i + u_i - w_i - 2w'_i)p_i\right)\left(\sum_{k=1}^{N} b_{0k}\right), \\
&= \sum_{i=1}^{N}\left[(2u'_i + u_i - w_i - 2w'_i)\left(b_{0i} - p_i\left(\sum_{k=1}^{N} b_{0k}\right)\right)\right], \\
&= 0.
\end{aligned}
\tag{23}
$$

### IV. MEAN DWELL TIME IN A SINGLE CYCLE

Let $T_k$ be the mean first passage time (MFPT) for the particle starting at mechanochemical state $k$ $(-N+1 \leq k \leq N-1)$ to reach the next mechanochemical cycle (backward or forward), i.e., to reach any of the states $i$ with $|i - k| \geq N$. Then the mean dwell time $T_{dwell}$ of the particle in a single mechanochemical cycle can be obtained as follows

$$T_{dwell} = \sum_{k=0}^{N-1} p_k T_k. \tag{24}$$

where $p_k$, $k = 0, \cdots, N-1$, is steady state probability of finding the particle in state $k$, which can be obtained by (13) and the normalizing condition (note, $p_0 = p_N$). The mean first passage time $T_k$ can be obtained by the method provided below. Due to the symmetry of $T_k$, we discuss only the case $k = 0$.

Let $\bar{T}_i$ be the mean first passage time of the particle starting at state $i$ $(-N+1 \leq i \leq N-1)$ to reach one of the absorbing boundaries $i = -N, -N-1, N, N+1$. Following [17], $\bar{T}_i$ is governed by

$$\bar{T}_i = \frac{1}{u''_i + w''_i}\left(1 + u_i \bar{T}_{i+1} + u'_i \bar{T}_{i+2} + w_i \bar{T}_{i-1} + w'_i \bar{T}_{i-2}\right), \tag{25}$$

for $-N+1 \leq i \leq N-1$, and with the absorbing boundary conditions

$$\bar{T}_i = 0, \quad \text{for } i = -N, -N-1, N, N+1. \tag{26}$$

For the sake of convenience, we write this equations as

$$H\vec{\bar{T}} = \vec{i}, \tag{27}$$

where $\vec{\bar{T}} = (\bar{T}_{-N+1}, \cdots, \bar{T}_{N-1})^T$, $\vec{i} = (1, \cdots, 1)^T$ are column vectors, while

$$\begin{aligned}
H_{i\,i-2} &= -w'_i, & \text{for } 3 \leq i \leq 2N-1, \\
H_{i\,i-1} &= -w_i, & \text{for } 2 \leq i \leq 2N-1, \\
H_{i\,i} &= u''_i + w''_i, & \text{for } 1 \leq i \leq 2N-1, \\
H_{i\,i+1} &= -u_i, & \text{for } 1 \leq i \leq 2N-2, \\
H_{i\,i+2} &= -u'_i, & \text{for } 1 \leq i \leq 2N-3; \\
H_{ij} &= 0, & \text{otherwise.}
\end{aligned} \tag{28}$$

Then the mean first passage time $T_0$ for the particle in state $k = 0$ to reach another mechanochemical cycle can be obtained by $T_0 = \bar{T}_0$. The mean first passage times $T_k$, for $k = 1, \cdots, N-1$, of the particle in state $k$ to reach another mechanochemical cycle (forward or backward), can be obtained by the same method. Finally, the mean dwell time of the particle in a single mechanochemical cycle can be obtained by formulation (24).

It can be checked that, if $u'_i = w'_i = 0$ for $1 \leq i \leq N$, i.e., the model reduces to the usual periodic one-dimensional hopping model, for which $T_{dwell} = T_0 = \cdots = T_{N-1}$.

## V. NUMERICAL RESULTS

To illustrate and verify the methods to calculate the mean velocity, effective diffusion constant, and mean dwell time, we applied them to a class of synthetic rotary molecular motors, which is recently devised by Feringa and coworkers [25, 29]. In fact, the model for such molecular motors is a special case of our modified one-dimensional hopping model with $N = 4$: see Fig. 2.

The parameters we used for numerical calculations are just those obtained experimentally in [29]. The transition rates are as follows (with unit s$^{-1}$)

$$u_1 = 27.4 \times 10^{-5}, \ u'_1 = 0.7 \times 10^{-5}, \quad u_2 = \frac{k_B T}{h} e^{-\frac{\Delta G_{23}}{RT}}, \ u'_2 = 0.35 \times 10^{-5},$$
$$u_3 = 20.6 \times 10^{-5}, \ u'_3 = 0.8 \times 10^{-5}, \quad u_4 = \frac{k_B T}{h} e^{-\frac{\Delta G_{41}}{RT}}, \ u'_4 = 0.55 \times 10^{-5},$$
$$w_2 = 2.1 \times 10^{-5}, \ w'_2 = 0.35 \times 10^{-5}, \quad w_1 = \frac{k_B T}{h} e^{-\frac{\Delta G_{14}}{RT}}, \ w'_1 = 0.7 \times 10^{-5}, \quad (29)$$
$$w_4 = 3.2 \times 10^{-5}, \ w'_4 = 0.55 \times 10^{-5}, \quad w_3 = \frac{k_B T}{h} e^{-\frac{\Delta G_{32}}{RT}}, \ w'_3 = 0.8 \times 10^{-5}.$$

Here $k_B = 1.38 \times 10^{-23}$ J/K is Boltzmann constant, $h = 6.626 \times 10^{-34}$ J·s is Planck constant, $R = 1.9872$ cal·K$^{-1}$·mol$^{-1}$ is gas constant, and the empirical values $\Delta G_{ij}$ are (kcal/mol)

$$\Delta G_{23} = 25.6, \quad \Delta G_{32} = 30.3, \quad \Delta G_{41} = 25.3, \quad \Delta G_{14} = 30. \quad (30)$$

Since the processes $1 \to 2$ and $3 \to 4$ are photochemical conversions, the transition rates $u_1, u_3, w_2, w_4$ and all the $u'_i, w'_i$ for $i = 1$ to 4 scale linearly with the light intensity $l$, for details see [29]. Illumination with $\lambda = 365$ nm corresponds to light intensity $l = 1$.

The numerical results of the mean velocity and effective diffusion constant as functions of temperature $T$ are plotted in Fig. 3. In the calculations, we assumed that the motor rotates $\pi/2$ in each of the four processes. The velocity $v = 1$ s$^{-1}$ means the motor complete one cycle per second, and similar meaning for the effective diffusion constant $D$. The figures in Fig. 3 are almost the same as the ones obtained in [29] (note, in their paper, different methods are used by Feringa and coworkers), which implies that our methods to get the mean velocity and effective diffusion constant of the modified one-dimensional hopping model are accurate to some extent.

To better understand the properties of this molecular motor, the numerical results of the mean velocity $V$ and effective diffusion constant $D$ as functions of light intensity $l$ are plotted in Fig. 4. One can easily see that, both $V$ and $D$ increase as $T$ and $l$. But, as pointed out in [29], the ratio of them, i.e., the Péclet number $Pe = V/D$ has a maximum as a function of $T$. Further calculations show that the Péclet number $Pe$ also has a maximum as a function of $l$: see Fig. 5.

Moreover, the maximum of $Pe$ satisfies

$$\max_{T,l} Pe(T,l) \approx \max_T Pe(T,l) \approx \max_l Pe(T,l) \approx 1.5 \tag{31}$$

So, for any given temperature $T$, we always can find an optimal value of light intensity $l_{opt}(T)$, at which the value of $Pe$ is maximum. But, roughly speaking, $l_{opt}(T)$ increases exponentially with $T$: see Fig. 6.

By the calculation results in Fig. 7, one finds that the mean dwell time $T_{dwell}$ decreases with $T$ and $l$, this is consistent with the corresponding results for the mean velocity: see Figs 3(a) and 4(a). However, there is no corresponding results about the mean dwell times $T_{dwell}$ in [29]. Instead, there is a dimensionless quantity, rotation excess, denoted by $r.e.$, is defined. In [20], it transpires that to some degree, a general $N$-state model can be simplified into a one-state model with *effective* forward and backward transition rates $u_{eff}$ and $w_{eff}$. For a one-state model, the probability flux is just $J = u_{eff} - w_{eff}$ while the mean dwell time in one single cycle is $1/(u_{eff} + w_{eff})$: see [16, 17]. So the effective transition rates $u_{eff}, w_{eff}$ can be obtained by requiring

$$u_{eff} - w_{eff} = J, \qquad u_{eff} + w_{eff} = \frac{1}{T_{dwell}}, \tag{32}$$

or, in other words, by

$$u_{eff} = \frac{1}{2}\left(\frac{1}{T_{dwell}} + J\right), \quad w_{eff} = \frac{1}{2}\left(\frac{1}{T_{dwell}} - J\right). \tag{33}$$

Physically, for the rotary molecular motor devised in [29], $u_{eff}$ and $w_{eff}$ denote the numbers of full forward and backward rotations per unit time, denoted by $\Omega_+$ and $\Omega_-$ in [29]. Therefore, the dimensionless rotational excess ($r.e.$) defined by Feringa and coworkers is given

$$r.e. = \frac{u_{eff} - w_{eff}}{u_{eff} + w_{eff}} = JT_{dwell}. \tag{34}$$

The calculation results plotted in Fig. 8 is almost the same as the ones obtained by Feringa and coworkers in [29], where a completely different and special method is used. This implies that, our results about the mean dwell times are accurate enough. In fact, the numerical results about $u_{eff}$ and $w_{eff}$ are also almost the same as the ones obtained in [29].

In conclusion, by applying our results about the mean velocity, effective diffusion constant, and mean dwell time, to the model of recently devised second generation rotary molecular motors, one can easily see that our results are accurate and valuable.

## VI. SUMMARY

In this paper, the modified periodic one-dimensional hopping model is discussed. In this model, the particle in state $i$ can not only jump forward to state $i+1$, or backward to state $i-1$, but also is allowed to jump to states $i+2$ or $i-2$ directly. One example is the model given by Feringa and coworkers for their second generation rotary molecular motors. Similar methods as in [10, 16] are used to get the mean velocity and effective diffusion constant, and the basic idea in [17] is employed to obtained the mean dwell time in one single mechanochemical cycle. The results are illustrated and verified by being applied to the model of Feringa. The method used in this paper is universal. It also can be employed for detailed studies of other types of one-dimensional hopping models, and consequently to theoretical analysis of varied kinds of biophysical and biochemical precesses, including the motion of many other synthetic molecular motors [30–35, 37–40].

### Acknowledgments

This work is funded by the National Natural Science Foundation of China (Grant No. 10701029).

---

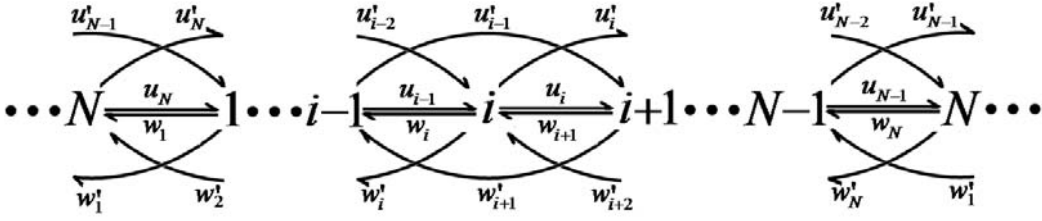

FIG. 1: Schematic depiction of our modified one-dimensional hopping model of period $N$, in which the particle in state $i$ can jump forward to state $i+1$ with rate $u_i$, to state $i+2$ with rate $u'_i$, or jump backward to state $i-1$ with rate $w_i$, to state $i-2$ with rate $w'_i$. The rates $u_i, u'_i, w_i, w'_i$ satisfy the periodicity conditions $u_{i+N} = u_i, u'_{i+N} = u'_i, w_{i+N} = w_i, w'_{i+N} = w_i$.

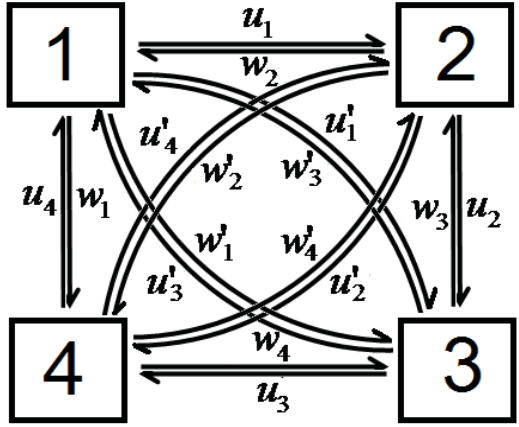

FIG. 2: Kinetic model of the synthetic rotary molecular motors recently devised by Feringa and coworkers [29]. The processes $1 \to 2$ and $3 \to 4$ correspond to photochemical conversions, while $2 \to 3$ and $4 \to 1$ correspond to subsequence thermal conversions. The rates $u_2, w_3, u_4, w_1$ depend on the environmental temperature $T$, and the transition rates $u_1, u_3, w_2, w_4$ and all the $u'_i, w'_i$ for $i = 1$ to $4$ scale with the light intensity $l$.

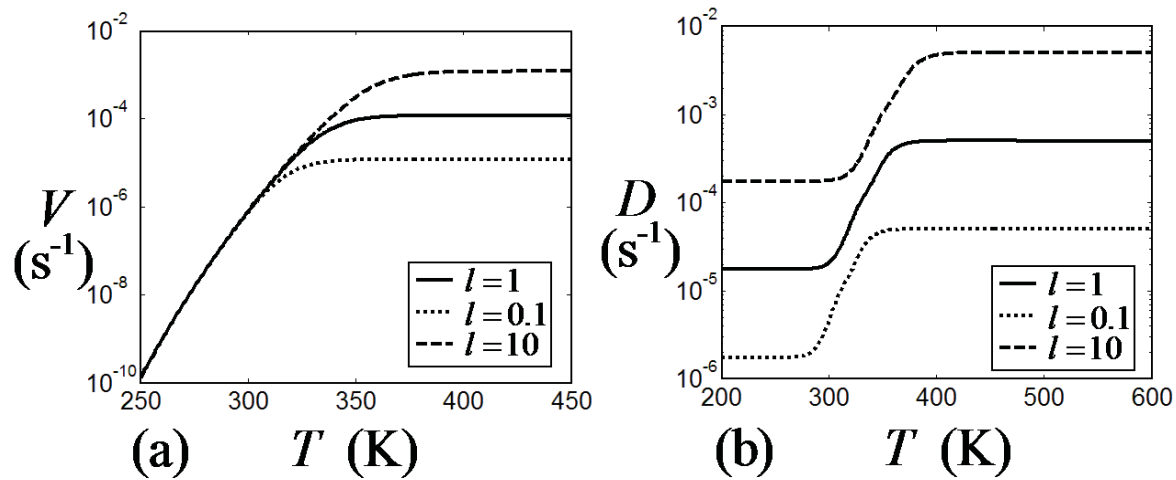

FIG. 3: The mean velocity (a) and effective diffusion constant (b) of the synthetic rotary molecular motor, as depicted in Fig. 2, as functions of temperature $T$.

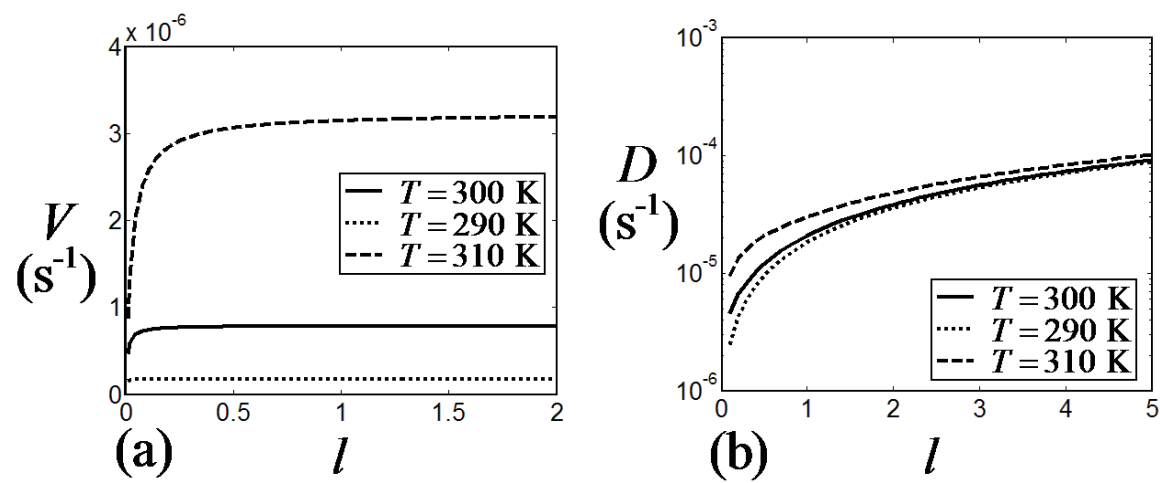

FIG. 4: The mean velocity (a) and effective diffusion constant (b) of the synthetic rotary molecular motor as functions of light intensity $l$.

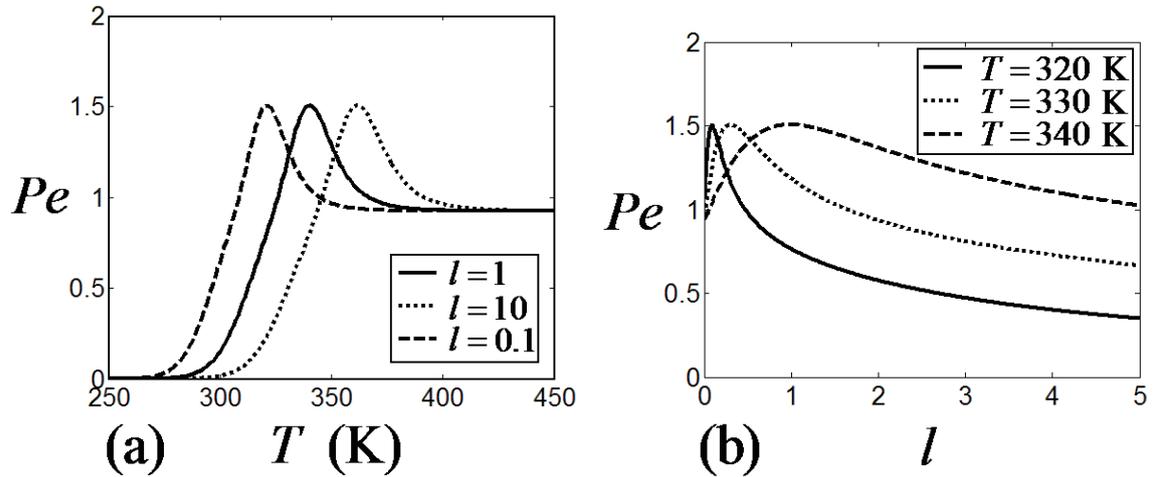

FIG. 5: The Péclet number $Pe = V/D$ has a maximum as functions of temperature $T$ and light intensity $l$.

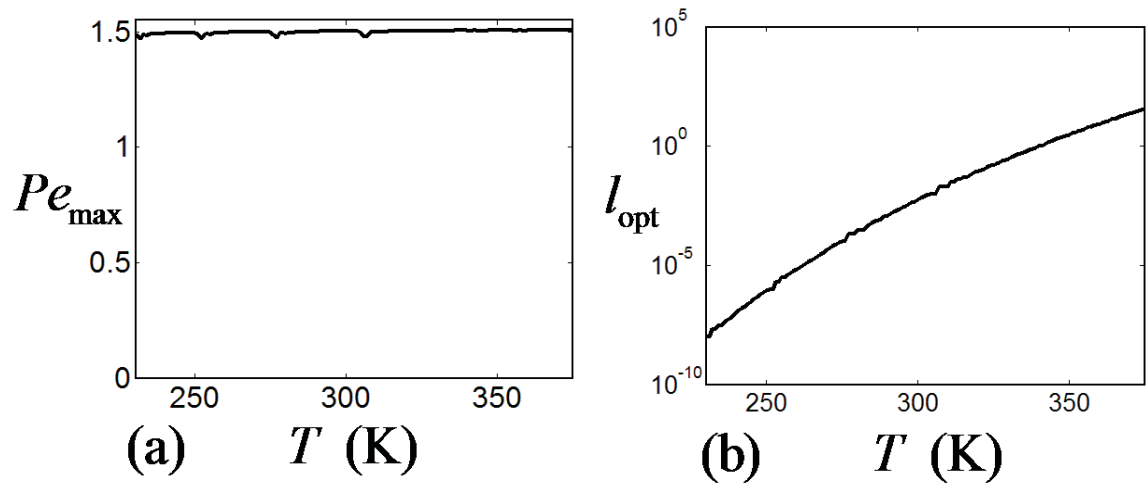

FIG. 6: (a) The maximum of the Péclet number as a function of temperature: $Pe_{max}(T) := \max_l Pe(T, l)$. (b) Roughly speaking, the corresponding values of light intensity $l_{opt}$, at which $Pe_{max}(T)$ is attended, increases exponentially with temperature $T$.

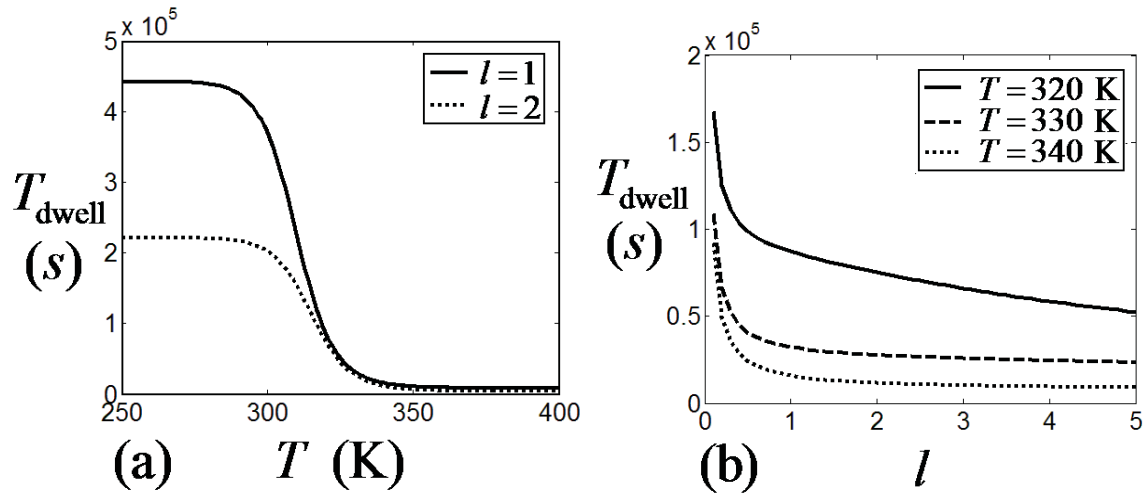

FIG. 7: Variation of the mean dwell time: $T_{dwell}$ decreases with $T$ and $l$.

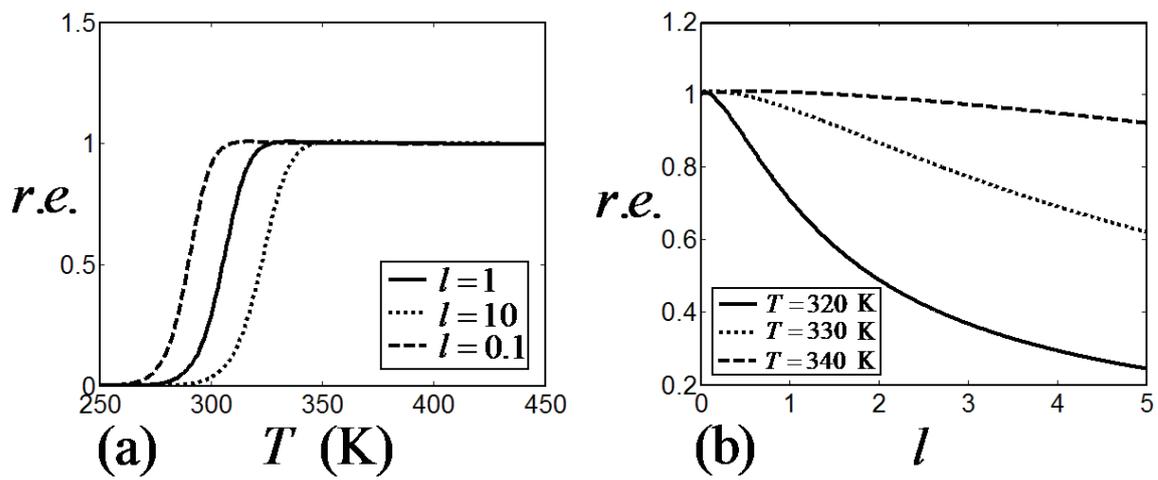

FIG. 8: The rotational excess $r.e. = JT_{dwell}$ as functions of temperature $T$ and light intensity $l$.